\documentstyle[aps,multicol,psfig]{revtex}

\newcommand{\be}{\begin{equation}}
\newcommand{\ee}{\end{equation}}
\newcommand{\bea}{\begin{eqnarray}}
\newcommand{\eea}{\end{eqnarray}}
\newcommand \ga{\raisebox{-.5ex}{$\stackrel{>}{\sim}$}}

\newcommand{\rx}{{\rm x}}
\newcommand{\ry}{{\rm y}}
\newcommand{\rz}{{\rm z}}
\begin{document}

\draft

\title{Anisotropic $J/\Psi$ suppression in nuclear collisions}

\author{Henning Heiselberg}

\address{NORDITA, Blegdamsvej 17, DK-2100 Copenhagen \O, Denmark}

\author{Raffaele Mattiello}

\address{Niels Bohr Institute, Blegdamsvej 17, DK-2100 Copenhagen \O, Denmark}

\maketitle

\begin{abstract}
The nuclear overlap zone in non-central relativistic heavy ion
collisions is azimuthally very asymmetric. By varying the angle
between the axes of deformation and the transverse direction of the
pair momenta, the suppression of 
$J/\Psi$ and $\Psi'$ will oscillate in a characteristic
way.  Whereas the average suppression is mostly sensitive to the early
and high density stages of the collision, the amplitude is 
more sensitive to the late stages.  This effect provides additional
information on the $J/\Psi$ suppression mechanisms such as direct
absorption on participating nucleons, comover absorption or formation
of a quark-gluon plasma.  The behavior of the average $J/\Psi$
suppression and its amplitude with centrality of the collisions is
discussed for SPS, RHIC and LHC energies with and without a phase transition.
\end{abstract}


\begin{multicols}{2}

\section{Introduction}

The observed suppression of $J/\Psi$ and $\Psi'$ in nuclear collisions
at SPS energies \cite{NA50} has created much interest, in particular
whether the ``anomalous'' suppression in central $Pb+Pb$ collisions
can be attributed to the formation of a quark-gluon plasma
\cite{Satz,Blaizot}.  It has been found (see, e.g., \cite{Kharzeev})
that in p+A reactions the suppression can be understood as Glauber
absorption of Charmonium states $\psi=J/\Psi,\Psi',\chi$, on
participant nucleons with $\sigma_{\Psi N}=7.3$~mb.  At SPS energies
also the $J/\Psi$ suppression in $S+A$ and peripheral $Pb+Pb$
collisions can be explained reasonably well but the model fails for
central Pb+Pb reactions.  Additional break up on {\it comovers} can,
however, also account for the observed strong suppression \cite{Vogt}
as is supported by microscopic approaches \cite{Capella,microsc}.
Understanding the early state dynamics in ultrarelativistic heavy ion
reactions is one of the major tasks in view of the upcoming
experiments at RHIC and LHC.

We will here suggest a new method to explore the $J/\Psi$ suppression
mechanism which exploits the anisotropy of the overlap zone in
semi-central collisions.  We will demonstrate how the azimuthal
distributions of charmonia depends on the initial asymmetry and can
give additional information on formation/dissociation processes, in
particular when and where they occur.  We find that the azimuthal
asymmetry is $\sim$5-10\% for semi-central Pb+Pb collisions at SPS
energies and more than doubles at RHIC and LHC energies. This is to be
compared to the present experimental uncertainty of 1-2\% in the
$J/\Psi$ minimum bias data of NA50 \cite{NA50}.  It is, however, less
than the typical elliptic flow $2v_2\simeq$~20\% at SPS \cite{NA49v2}.
Measuring the reaction plane in coincidence with $J/\Psi$'s is
therefore a promising method which provides new insight in $J/\Psi$
absorption mechanisms and possible phase transitions.

The two important mechanisms for $J/\Psi$ suppression, comover and
Glauber absortion, are discussed in Sec. II and III respectively with
emphasis on the azimuthal asymmetries.  Result for SPS, RHIC and LHC
energies are presented in Sec. IV, effects of a phase transition is
discussed in Sec. V, and finally a summary is given.

\section{Asymmetries in comover absorption}

We describe the evolution of the charmonium distribution 
by the Boltzmann equation
\bea
 &&\left( \frac{\partial}{\partial t} + {\bf v}_p\cdot \nabla
\right) n_\psi 
 = - \int\frac{d^3p_c}{(2\pi)^3} \, v_{\psi c}\sigma_{\psi c}\,
  n_\psi n_c
 \,, \label{BE}
\eea
where $n_i(r_i,p_i,t)$ are the particle distribution functions,
$\sigma_{\psi c} $ is the comover absorption cross section for scatterings
of charmonia $\psi$ with a comover $c$
and $v_{\psi c}$ their relative velocities.

Let us first consider peripheral nuclear collisions where the overlap
zone and the produced particle densities are small. Collisions are
then rare and their effects can be calculated to first order with the
free streaming distribution functions in the collision term of the
Boltzmann equation Assuming Bjorken scaling along the beam or
$\rz$-axis, $v_\rz=\rz/t$. The free streaming distribution can at
later times be approximated by \cite{Baym,v2}
\bea
  n(x,p)\simeq\frac{(2\pi)^3}{\tau m_\perp}
\delta(y-\eta)\frac{dN}{dyd^2p_\perp}\,
   S_\perp({\bf r}-{\bf v}(\tau-\tau_0)) \,. \label{fs}
\eea
Here, ${\bf r}=(\rx,\ry)$ is the transverse radius and
${\bf v}=(v_x,v_y)$ the transverse velocity of the scatterers or comovers.
As usual, $\eta=(1/2)\log((t+\rz)/(t-\rz))$ is the space-time rapidity
and $\tau=\sqrt{t^2-\rz^2}$ is the invariant time.
$\tau_0$ is the formation time for comovers.

The transverse particle density distribution, $S_\perp(\rx,\ry)$, is for
convenience approximated by simple gaussians
\bea
S_\perp(\rx,\ry)  = \frac{1}{2\pi R_xR_y}
  \exp\left( - \frac{\rx^2}{2R_x^2} -  \frac{\ry^2}{2R_y^2}\right)
   \,. \label{gaussian}
\eea
The rms radii of the overlap zones $R_x$ and $R_y$ 
at collision are determined from nuclear thickness functions \cite{v2}.
The $\rx$-direction is chosen along the impact parameter ${\bf
b}$ and $(\rx,\rz)$ constitutes the {\it reaction plane}
with $\ry$ perpendicular to the reaction plane.
Replacing the gaussian transverse particle distribution by a constant density
inside an ellipse with axes $R_{x,y}$ in Eq. (\ref{gaussian})
does not change results by much. The only relevant scale is
$v_{x,y}\tau/R_{x,y}$ which contains the direction of ${\bf p}_\perp$
and the transverse anisotropy.

The integration over transverse coordinates and rapidity is straight
forward (see \cite{v2} for details).  The scatterer momenta can also
be summed over and in the following we will by $\langle .. \rangle$
indicate averaging over scatterer momenta $p_c$.  Only an integral
over proper time is left. With $N_i=dN_i/dp_\perp dy$ we obtain for
the comover absorption
\bea
 \frac{dN_\psi }{d\tau} &=& - N_\psi  \tilde{\sigma}_\psi  \, 
\frac{1}{\tau}
 \exp\left[-(\frac{\tau}{\tau_0}-1)^2 \xi^2  \right] \,, \label{J1}
\eea
where we have defined the dimensionless quantity
\bea
 \xi\equiv(\tau_0/2)\sqrt{\langle v_{x,\psi c}^2/R_x^2 + v_{y,\psi c}^2/R_y^2}
\eea
and the cross section times a transverse
particle density, i.e. an effective ``opacity'', as
\bea
  \tilde{\sigma}_\psi  \equiv\sum_c
  \frac{dN_c}{dy} \frac{\langle v_{\psi c}\sigma_{\psi c}\rangle}{4\pi R_xR_y}
   \,, \label{s}
\eea

 In the participant or ``wounded nucleon'' model, which seems
to describe the gross properties of relativistic nuclear collisions
reasonably, both the rapidity density $dN/dy$ and the transverse
area $R_xR_y$ scales approximately with $E_T$, except for very
peripheral collisions (see, e.g., \cite{Capella,v2}).  Therefore the
opacity $\tilde{\sigma}$ of Eq. (\ref{s}) is approximately {\it constant}
for central and semi-central collisions but decreases for very
peripheral collisions. The almost constant opacity is different from
the comover model of \cite{Vogt} which assumes densities to scale
linearly with transverse energy.

The dependence on the azimuthal angle $\phi$ between the reaction
plane and the charmonium 
transverse velocity $(v_x,v_y)_\psi =v_\psi (\cos\phi,\sin\phi)$, (see Fig. 1)
is coupled to the source deformation
\bea
  \delta \equiv \frac{R_y^2-R_x^2}{R_y^2+R_x^2} \,. \label{delta}
\eea
Defining the mean radius $R^2\equiv(R_x^2+R_y^2)/2$, we obtain
\bea
 \xi^2 &\simeq& \frac{\langle v_{\psi c}^2\rangle\tau_0^2}{4R^2}  
\left(1+\delta
\frac{v_\psi^2}{\langle v_{\psi c}^2\rangle} \cos(2\phi)\right) 
\,.\label{x}
\eea
Here, we have expanded assuming
$\delta(v_\psi^2/\langle v_{\psi c}^2\rangle)\ll 1$, which is valid 
for charmonia with $p_\perp\simeq 1$ GeV.

The deformation can be approximated from the full transverse extent of
the initial overlap of two nuclei with radius $R_A$ colliding with
impact parameter $b$ (see Fig. 1). They are simply: $R_x=R_A-b/2$ and
$R_y=\sqrt{R_A^2-b^2/4}$, and the corresponding deformation is
$\delta=b/2R_A$.  Taking the rms radii of the nuclear overlap zone
weighted with longitudinal thicknesses gives almost the same and may
be related to the elliptic flow and HBT radii \cite{v2}.

For peripheral collisions, where the comover absorption is small,
we obtain by integrating Eq. (\ref{J1}) from the comover formation
time $\tau_0$ to infinity
\bea 
 -\frac{\Delta N_\psi }{N_\psi }&=& \tilde{\sigma}_\psi  F(\xi) \nonumber\\
 &\simeq& \tilde{\sigma}_\psi  
  F(\frac{\langle v_{\psi c}\rangle\tau_0}{2R}) - 
 \frac{\tilde{\sigma}_\psi \delta}{2}
  \frac{v_\psi^2}{\langle v_{\psi c}^2\rangle} \cos(2\phi) 
  \,,  \label{J2} 
\eea
where  we have introduced the function ($C=0.557..$)
\bea
  F(\xi) \equiv \int_0^\infty \frac{e^{-x'^2}dx'}{\xi+x'}
 =  \left\{ \begin{array}{lrl}
   \log(\frac{1}{\xi})-\frac{C}{2} & , & \xi\ll 1 \\
   \sqrt{\pi}/2\xi   & , & \xi\gg 1 \end{array} \right\} 
   \,.   \label{F}
\eea
With typical relative velocities and formation times we find
$\langle v_{\psi c}\rangle\tau_0\ll R$ 
except for the very peripheral collisions.
>From Eq. (\ref{x}) thus $\xi\ll 1$ for central and semi-central collisions.

The result of Eq. (\ref{J2}) for comover absorption gives the average
suppression (see Fig. 2) as well as a characteristic oscillation with
the azimuthal angle between reaction plane and the transverse momentum
of the charmonia (see also Fig. 3).
It is important to observe that 
the average and the amplitude of the absorption probes different
space-times of the collision and therefore different physics:
{\it The average absorption occurs at early times}
because densities decrease inversely with proper time which leads to
the logarithmic dependence of the initial time $\tau_0$ as is also
found in the comover model of \cite{Vogt}. Contrarily, {\it the
amplitude of the oscillation is more sensitive to later times} of order
$\tau\simeq 2R/v_\psi$ due to the extra factor $(\tau-\tau_0)^2$ that
appears in (\ref{J1}) when expanding for small deformations as in
Eq. (\ref{J2}). The physics behind this late time effect on the
amplitude, is that the particle has to travel a distance $\sim R$
before the deformation of the source leads to more or less
absorption. Likewise the amplitudes decrease with the transverse
velocity as $v_\psi^2$ because the slow charmonia do not have the time
travel the distance $R$ before the comovers have expanded on a time
scale $R/v_c$.

For more central collisions the free streaming assumption breaks down
as interactions between comovers drive the system towards thermal
equilibrium and hydrodynamic expansion. Also the absorption is a
significant fraction and the linear expansion of Eq. (\ref{J2}) breaks
down. For simplicity, we shall assume that the 
free streaming result of Eq. (\ref{J1})
is approximately valid for the comovers in central collisions as well. 
The heavy
charmonia do not thermalize but rather break up when interacting and therefore
the distribution function for the remaining charmonia is free streaming
to a good approximation.
By taking proper time derivative of Eq.  (\ref{J1}) and
treat $N_\psi $ as a decreasing function with time, we obtain by
integrating over time that
\bea
 \frac{N_\psi}{N_\psi(\tau_0)}&\simeq&e^{-\tilde{\sigma}_\psi F(l)}\label{J3}\\
 &\simeq&  \left(\frac{\langle 
 v_{\psi c}\rangle\tau_0}{e^{-C/2}2R}\right)^{\tilde{\sigma}_\psi }
 \left(1+\frac{\tilde{\sigma}_\psi \delta}{2}
 \frac{v_\psi^2}{\langle v_{\psi c}^2\rangle}  \cos(2\phi) \right)
   \,, \label{Jexp}
\eea
where the latter expansion is valid for central and semi-central
collisions where $\langle v_{\psi c}\rangle\tau_0/R$ and $\delta$ are
small.  $N_\psi (\tau_0)$ is the number of $J/\Psi$'s at time $\tau_0$
where comover absorption sets in. 

In the wounded nucleon model also the transverse energy $E_T$ scales
with the transverse area, i.e., $R_xR_y\simeq R^2\propto E_T$.
Consequently, the $J/\Psi$ suppression decreases with centrality as a
power law, $J/\Psi\propto E_T^{-\tilde{\sigma}/2}$. 
It differs from the exponential fall-off of the
comover model of \cite{Vogt} that results from assuming that number
densities and thus effective opacities scales linearly with $E_T$.

The transverse momentum, rapidity and centrality dependence of the
amplitude in Eq. (\ref{Jexp}) differs from the average suppression.
The amplitude is increase proportionally to $dN/dy$ and $p_\perp^2$
but decreases as $\delta$ almost linearly with centrality, $E_T$.  The
relative amplitudes for $J/\Psi,\Psi',\chi$, etc.  are proportional to
the absorption cross section and should therefore scale with these
when corrected for feeddown whereas the DY background should not
oscillate.

We note that the prefactor to $\cos\phi$ is exactly the elliptic flow
parameter ($2v_2$) if the absorption cross section is replaced by the
momentum transport cross section, $\sigma_{tr}$ \cite{v2}.  The
elliptic flow of pions and proton $v_2\sim 0.1$ for pions and protons
could be described with $\sigma_{tr}\simeq 10$~mb.  Therefore, the
magnitude of the charmonium absorption amplitude is approximately
$\sim 2v_2 \sigma_{\psi c}/\sigma_{tr}\simeq$5\% depending on rapidity
and transverse momentum.

\section{Glauber absorption}

Glauber absorption of charmonium on nucleons 
is believed to be more important at SPS energies than comover absorption. 
The dissociation cross sections on nucleons and 
secondaries depend on various model assuptions.
Typically, for nuclear absorption 
one takes $\sigma_{\psi N}\simeq$3-5~mb and assumes 
that comovers scatter with cross sections 
varying from $\sigma_{\psi c}\simeq 1-5$mb \cite{Capella,Vogt,microsc}.
The relative amount of Glauber versus comover absorption is, however,
unresolved at present since 
it is also possible to describe the charmonium by Glauber absorption
--at least for p+A and S+U reactions-- 
taking $\sigma_{\psi N}\simeq$7~mb \cite{Kharzeev}.
Furthermore, it has been pointed out that 
feed-down effects from $\Psi'$ and $\chi$-decays into J/$\Psi$ \cite{Gerland}
or formation times \cite{Hufner,microsc} can play a major role.

As we shall concentrate on azimuthal asymmetries, we choose
the standard model for Glauber absorption
\bea
 G_{AB}(b) &=& {\cal N} \int d^3r\,d^3r' \rho_A(r)\rho_B(r')
   \delta^2({\bf b}-{\bf r}_\perp+{\bf r}_\perp') \nonumber\\
  && \times \exp\left(-L_A(r)-L_B(r')\right) \,, \label{Glauber}
\eea
where $\rho_{A,B}$ are the nuclear densities,
$L_A(r)=\sigma_{\psi N}\int_z^\infty dz_A\rho_A({\bf r}_\perp,z_A)$,
and analogously for $L_B(r')$. The normalization constant ${\cal N}$
is chosen such that $G_{AB}(b)=1$ when $L_A=L_B=0$.
For simplicity we use the Glauber absorption 
of $J/\Psi$ with $\sigma_{J/\Psi N}=3.6$~mb as in \cite{Gerland} and
no formation time.

The straight line geometry implied by the Glauber model for absorption
does not contain any azimuthal asymmetry, i.e., it is independent of
$\phi$. In the NA50 experiment the $J/\Psi$'s are collected at
pseudorapidies $\eta=2.8-4$ corresponding to angles of 2-7 degrees
with respect to the beam axis.  The $J/\Psi$'s thus traverses a
transverse distance $r_\perp\simeq
Rv_\perp\cosh(\eta)\simeq0.2-0.9$~fm inside the nuclei and their
absorption might be affected by the density distribution variation
transversely.  However, the net absorption from opposite direction
depends on the curvature of and not the gradient of the transverse
density distribution.  Therefore, the resulting azimuthal dependence
of Glauber absorption is $\sim (r_\perp/R_x)^2-(r_\perp/R_y)^2\sim
\delta (r_\perp/\bar{R})^2$.  It amounts to a few percent at
midrapidity but is negligible at the forward rapidities. 

Momentum broadening affects the initial charmonium production rates
and is another possible mechanism which could introduce azimuthally
dependence in Glauber absorption.  Momentum broadening has been found
in pA and BA collision at relativistic energies and depends on the
number of NN collisions and therefore also the nuclear thickness
function.  The momentum broadening could be sensitive to the varying
density profile depending on the range of the interaction, $r_i$.  As
above the resulting azimuthal dependence of Glauber absorption is
$\sim (r_i/R_x)^2-(r_i/R_y)^2$. Furthermore, it has to be multiplied
with the relative momentum broadening. With $r_i\sim 1$~fm we estimate
that this effect is also small.  Since DY pairs should be affected by
momentum broadening of the cascading nucleons like Glauber apsorption
but not by comover absorption, it is possible to discriminate between
the azimuthal dependence of Glauber and comover absorption by
measuring the amplitude for Drell-Yan (DY) production.

These rough estimates clearly deserve more study in detailed
microscopic models which, however, is outside the scope of this work.
We emphasize that the time scales and distances involved for
the initial Glauber scatterings are very short. Causality limits
particles to probe transverse distance scales shorter than $\sim R_z
/\gamma$, where $\gamma$ is the Lorentz contraction factor. Such short
distances are to be compared to the transverse radii, $R_{x,y}$, over
which densities vary azimuthally.

Predictions for RHIC and LHC energies depend on the charmonium
absorption cross section at these energies. If formation times are
$\tau_\psi\ga R/\gamma$, the Glauber absorption might dissappear or be
enhanced at higher energies due to time dilation \cite{Hufner} and
different production mechanisms as e.g. suggested by the color-singlet
or the color-octet model.  However, pA Fermilab data on $J/\Psi$
production \cite{Fermilab} indicates that the absorption cross section
does not change going from 200 to 800 GeV lab energy.  For the
estimates presented below we therefore assume that the Glauber
absorption is unchanged going from SPS to RHIC and LHC
energies. Furthermore, the minor azimuthal dependence of Glauber
absorption will be ignored.

\section{Estimates for SPS, RHIC and LHC}

First we estimate the parameters in play. The unknown ones are the
comover absorption cross section, $\sigma_{\psi c}$, and the formation
time $\tau_0$. The other parameters and their dependence on impact
parameter can be estimated from experiments and results in an almost
constant opacity as explained above.  We give an estimate for central
$Pb+Pb$ collisions at SPS energies. Here, the rms radius is $R^2\simeq
10$fm$^2$, and the total rapidity density is $dN^{tot}/dy\simeq
500$. One can assume \cite{Vogt} that only heavy resonances as
$\rho,\omega,\phi$, etc. have sufficient energy to destroy the
charmonia and take $dN_c/dy\simeq 0.5dN^{tot}/dy$. The average
relative velocity for charmonia with transverse momentum $\sim1$ GeV/c
is $v_{\psi c}\simeq 0.6c$. A comover absorption cross section of
order $\sigma_{\psi c}=(2/3)\sigma_{\psi N}\simeq 2.4$~mb is typically
assumed, and so we obtain the opacity $\tilde{\sigma}\simeq 0.3$ for
central and semi-central collisions.  
Its dependence on centrality for
peripheral collisions is calculated in the wounded nucleon model
\cite{v2}. 

At RHIC and LHC energies the rapidity density is expected
to increase with collision energy and we shall simply assume that they
double at RHIC energies and triple at LHC energies. Consequently, the
opacity is $\tilde{\sigma}=$0.3, 0.6, and 0.9 at SPS, RHIC and LHC
energies respectively.  In Fig. 2. we show the corresponding comover
suppression of the $J/\Psi$ from Eq. (\ref{J3}-\ref{Jexp}) assuming
$\tau_0=1$fm/c and taking the same standard Glauber absorption from
Eq. (\ref{Glauber}) with $\sigma_{\Psi N}=3.6$~mb at all three energies.

The modulation of the $J/\Psi$-suppression with angle between
$p_\perp$ and reaction plane is shown in Fig. 3 for five centralities
or $E_T$ corresponding to impact parameter $b/R\simeq 2,1.5,1,0.5,0$.
For a prediction at RHIC we take the same Glauber absorption as in
Fig. (2) as well as the SPS values for $v_\psi \simeq 0.3c$ and
$v_{\psi c}\simeq 0.6c$. However, $\tilde{\sigma}\simeq 0.6$ is
employed for RHIC energies due to the larger rapidity densities.  The
relative amplitude is given by Eq. (\ref{Jexp}) and decreases with
centrality as the deformation. For near-central collisions the source
expands significantly and the deformation may be smaller than that at
initial overlap, $\delta\simeq b/2R$. The magnitude of the amplitude
is up to $\pm 7\%$ for semi-peripheral events. As the relative
amplitude scales with the rapidity density, it is smaller by a
factor of two at SPS energies but larger at LHC energies.

The relative amplitudes should be measurable by techniques similar to
those used for extracting elliptic flow \cite{NA49v2}. By taking the
ratio of the amplitude to the average $J/\Psi$ or using a minimum bias
analysis, the largest statistical uncertainty (the DY background) is
removed.  The NA50 minimum bias data has a relative error bar of
1-2\%, i.e., an order of magnitude less than the variation in the
$J/\Psi$ absorption,
$[N_\psi(\phi=0)-N_\psi(\phi=\pi/2)]/\bar{N}_\psi$, which is twice the
amplitude in Eq. (\ref{Jexp}).

It would also be most interesting to measure and compare the amplitude for
the various charmonium states $J/\Psi$, $\Psi'$, $\chi$ as well as for the DY
background. The $\Psi'$ is expected to have a larger dissociation cross 
section and the amplitude should be correspondingly 
larger. The amplitude for
the DY background should vanish as discussed in the previous section.

\section{Effects of a phase transition}

The anomalous $J/\Psi$ suppression measured by NA50 is possible due to
the formation of a plasma state leading to color screening of
$c\bar{c}$ pairs \cite{Satz,Blaizot}. As the plasma is created at high
energy densities present at early times it would increase the average
suppression.  However, the amplitude would be less affected since it
is caused by late time absorption.  We can mimic this situation by
employing a larger absorption cross section at early times than at
late times or in terms of Eq. (\ref{Jexp}) a larger $\tilde{\sigma}$
in the average power law suppression than in the prefactor to the
amplitude. Both quantities can be measured as function of centrality
and it would be most interesting to follow their behavior for the
$J/\Psi$ and $\Psi'$ suppression as well as the subthreshold
enhancement - in particular in the regions of anomalous $J/\Psi$ and
$\Psi'$ and suppression.

If the charmonium absorption cross sections are strongly energy
dependent \cite{Kharzeev}, the average cross section $\langle v_{\psi
c}\sigma_{\psi c}\rangle$ may also be larger at earlier times since
the relative energies are larger here.  In the above calculations we
ignored relative longitudinal velocities by assuming the
$\delta(y-\eta)$ factor in the distribution. This approximation breaks
down at early times and one should therefore take an effective cross
section at larger collision energy in Eq. (\ref{Jexp}) at early
times. However, whereas the energy dependent cross sections should
lead to a smooth variation of the average $J/\Psi$-suppression as well
as its amplitude, the plasma leads to an anomalous and sudden
suppression at a certain centrality or $E_T$ \cite{Blaizot,Kharzeev}.

\section{Summary}

In summary, we have investigated suppression of charmonia in
non-central collisions and find a characteristic oscillation with
azimuthal angle between transverse momentum and reaction plane due to
comover absorption.  Glauber absorption was estimated to contribute
less to the amplitude because causality strongly constrains the
transverse distances probed and the resulting azimuthal dependence.
Therefore, the azimuthal variation may be exploited to gain more
information on the suppression mechanisms.

Estimates for comover absorption predicts an azimuthal variation of
order 5-20\% in the number of $J/\Psi$ depending on centrality, size
and energy of the colliding nuclei. Such an oscillation should be
detectable (if the reaction plane can be determined event-by-event) as
it is not sensitive to the statistical uncertainty of the DY
background. In comparison, the current minimum bias data of NA50 has
error bars of 1-2\%.

The amplitude has a strong dependence on rapidity and transverse
momentum of the charmonia and on centrality that is different from the
average suppression.  The amplitude is proportional to $dN/dy$ and
$p_\perp^2$ and decreases almost linearly with centrality $E_T$.  The
relative amplitudes for $J/\Psi$, $\psi'$ and $\chi$ are proportional
to the absorption cross section and should therefore scale with these
when corrected for feeddown.  The DY background should, however, not
oscillate.

The average absorption and its amplitude are more sensitive to early
and later times respectively. This distinction may further be exploited to
obtain additional information on the absorption mechanism and whether
an anomalous suppression is due to a phase transition.



\end{multicols}


\begin{figure}
\centerline{\psfig{figure=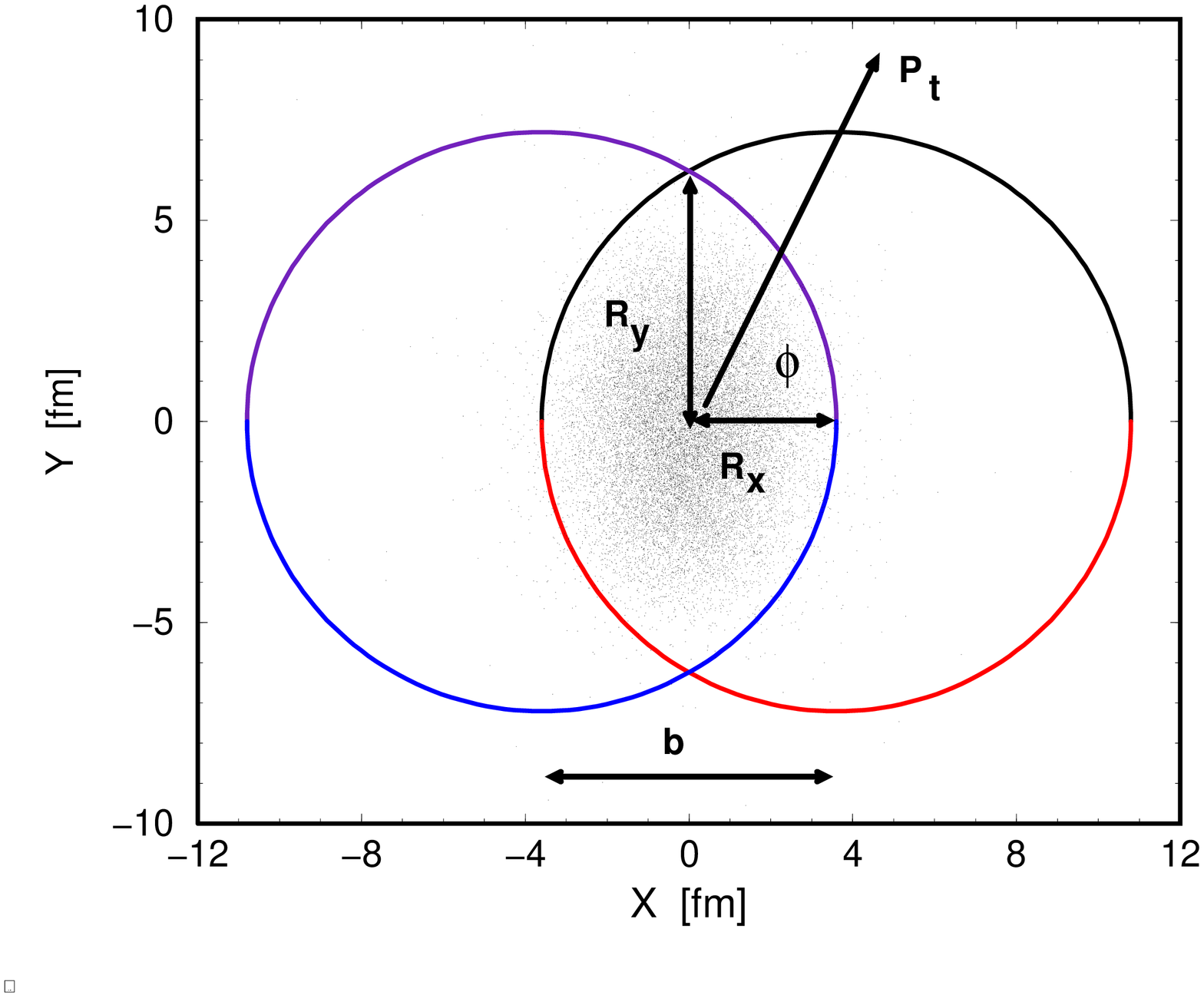,width=12cm,height=10cm,angle=-90}}
\vspace{0.2cm}
\caption{Reaction plane of a semi-central $Pb+Pb$ collision with impact 
parameter $b=R_{Pb}\simeq8$fm. The overlap
zone is deformed with $R_x\le R_y$. The reaction plane $(\rx,\rz)$ is
rotated by the angle $\phi$ with respect to the (outward)
transverse particle momentum $p_\perp$. The scatter plot shows charmonium 
dissociation points in a microscopic cascade simulation 
\protect\cite{Mattiello} at SPS energies.
\label{geofig}  }
\end{figure}

\vspace{-1.5cm}
\begin{figure}
\centerline{\psfig{figure=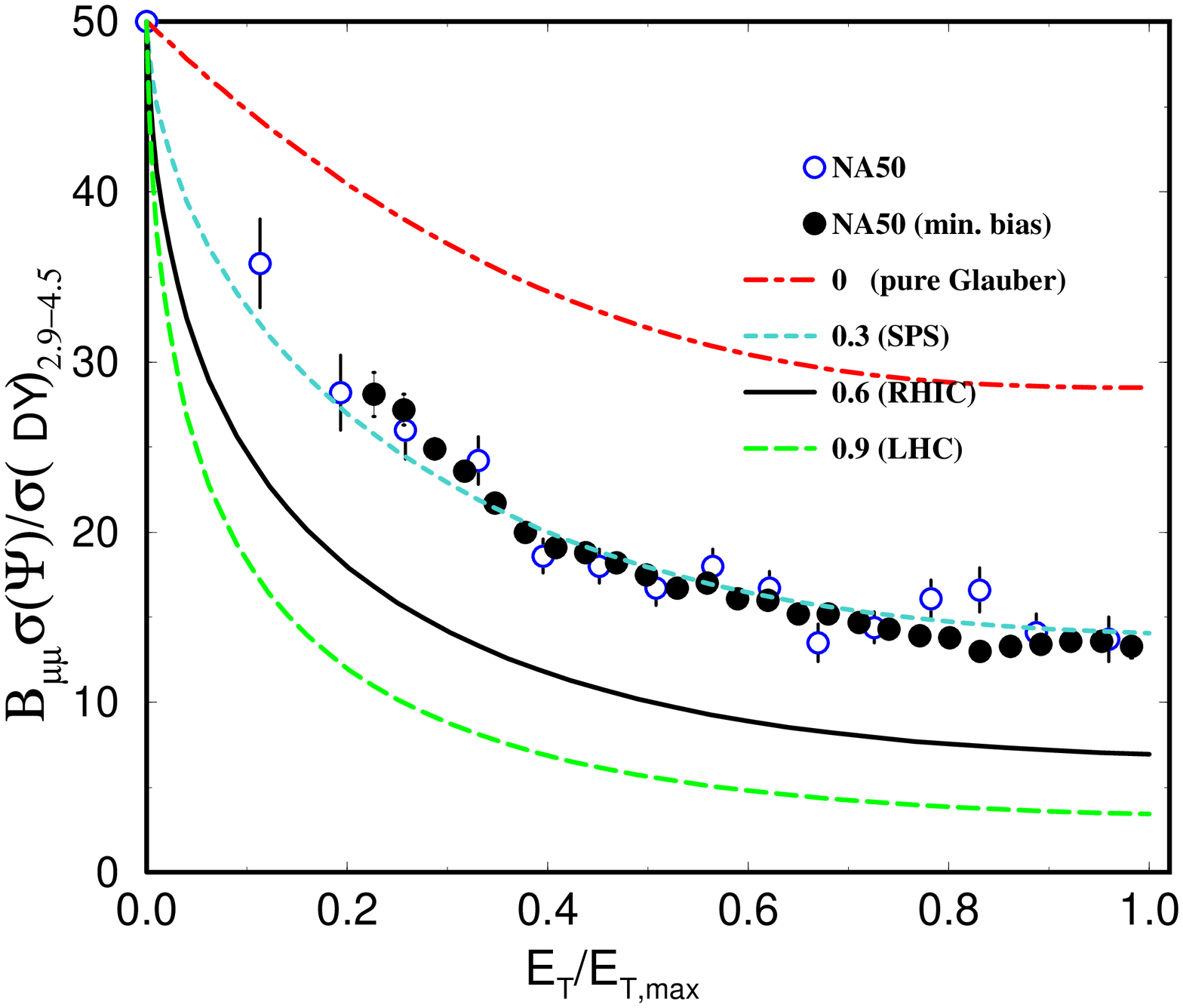,width=12cm,height=10cm,angle=-90}}
\vspace{0.5cm}
\caption{Comover and Glauber suppression of $J/\Psi$,
Eqs. (\ref{J3}-\ref{Glauber}),
shown by curves as function of transverse energy 
$E_T$. The opacities are: $\tilde{\sigma}=0.0, 0.3, 0.6, 0.9$ corresponding
approximately to pure Glauber absorption, SPS, RHIC and LHC energies 
(see text for details). Also shown
is NA50 data \protect\cite{NA50} normalized to DY and minimum bias
(open and filled circles respectively).}
\end{figure}

\vspace{-1.5cm}
\begin{figure}
\centerline{\psfig{figure=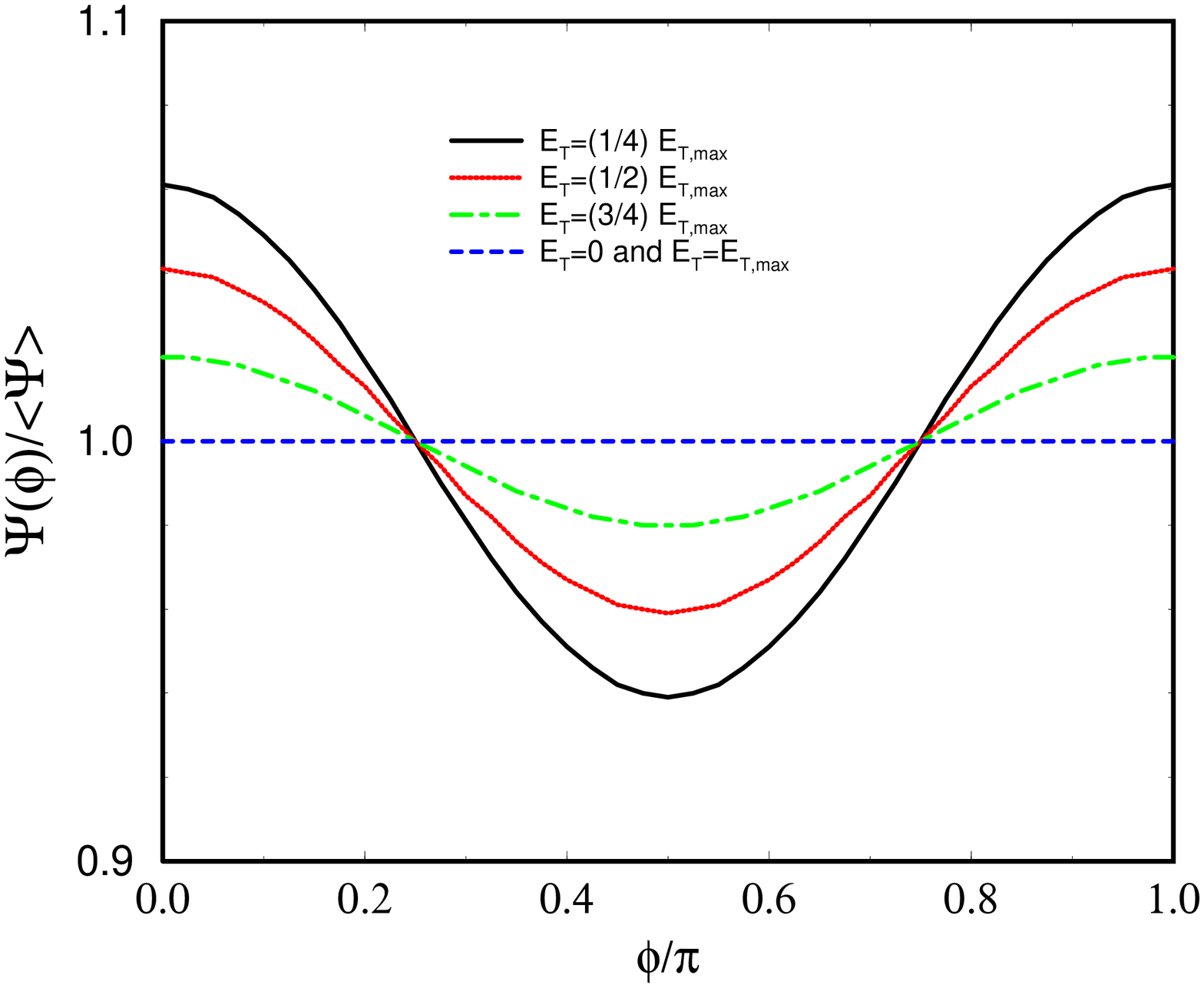,width=12cm,height=10cm,angle=-90}}
\vspace{0.5cm}
\caption{$J/\Psi$ suppression as function of azimuthal 
angle $\phi$ normalized to its average 
for various centralities from Eq. (\ref{Jexp}). The RHIC parameters
$\tilde{\sigma}=0.6$ are employed (see text).}
\end{figure}


\begin{thebibliography} {999}

\bibitem{NA50} NA50 Collaboration, M.C. Abreu et al.,
{\it Phys. Lett.} {\bf B410}, 327 (1997); {\it ibid} 337;
CERN-EP/99-13, to appear in Phys. Lett. B.
\bibitem{Satz} T. Matsui and H. Satz, Phys. Lett. 178B, 416 (1991)
\bibitem{Blaizot} J.P. Blaizot and J.Y. Ollitrault, {\it Phys. Rev. Lett.}
{\bf 77}, 1703 (1996).
\bibitem{Kharzeev} D. Kharzeev, Nucl. Phys. 638A (1998), 279c and 
Refs. therein
\bibitem{Vogt} S. Gavin and R. Vogt,  Phys. Lett. 207B, 257 (1988);
 Phys. Lett. 345B, 104 (1988); {\it Phys. Rev. Lett.} {\bf 78}, 1006 (1997);
\bibitem{Capella} N. Armesto, A.Capella, E. Ferreiro, {\it Phys. Rev.}
{\bf C59}, 395 (1999).
\bibitem{microsc} 
C.Y. Wong, {\it Nucl. Phys.} {\bf A610}, 434c (1996);
W. Cassing, E. Bratovskaya, Nucl. Phys. 623A, 570 (1997);
W. Cassing, C.M. Ko, Phys. Lett. 387B, 691 (1996);
C. Spieles et al., Eur. Phys. J. C5, 349 (1998), 
\bibitem{Gerland} L. Gerland et al., Phys. Rev. Lett. 81, 762 (1998)
hep-ph/9810486.
\bibitem{NA49v2} H. Appelsh\"{a}user et al. (NA49 collaboration), 
    {\em Phys. Rev. Lett.} {\bf 80}, 4136 (1998).
\bibitem{Baym} G. Baym, {\it Phys. Lett.} {\bf 138B}, 18 (1984).
  H. Heiselberg and X.-N. Wang, {\it Phys. Rev.} {\bf C53}, 1892 (1996).
\bibitem{v2} H. Heiselberg and A. Levy, {\it Phys. Rev.} C (1999),
nucl-th/9812034.
\bibitem{Hufner} J. Hufner and B.Z. Kopeliovich, {\it Phys. Lett.}
{\bf B445}, 223 (1998). 
\bibitem{Fermilab} A. Sansoni, (CDF coll.), {\it Nucl. Phys.} {\bf A610},
         373c (1996).
\bibitem{Mattiello} R. Mattiello and J.P. Bondorf, proc. of the RIKEN
Workshop, 'High Density Matter in AGS, SPS and
RHIC Collisions', July 11, 1998; and in preparation.

\end{thebibliography}
\end{document}